\documentclass[10pt, conference, compsocconf]{IEEEtran}

\usepackage{cite}
\usepackage{graphicx}
\ifCLASSINFOpdf
\else
\fi
\usepackage[cmex10]{amsmath}
\usepackage{multirow}
\usepackage{algorithmic}
\usepackage{array}

\usepackage{mdwmath}
\usepackage{mdwtab}
\usepackage[tight,footnotesize]{subfigure}
\usepackage[font=footnotesize]{subfig}
\usepackage{url}
\begin{document}
%
\title{The Effect of Pets on Happiness: \\
	A Data-Driven Approach via Large-Scale Social Media  }
    
\author{\IEEEauthorblockN{Yuchen Wu}
	\IEEEauthorblockA{Electrical and Computer Engineering\\
		University of Rochester\\
		Rochester, NY 14627\\
		ywu55@ece.rochester.edu}
\and
\IEEEauthorblockN{Jianbo Yuan, Quanzeng You and Jiebo Luo}
\IEEEauthorblockA{Department of Computer Science\\
University of Rochester\\
Rochester, NY 14627\\
{jyuan10, qyou, jluo}@cs.rochester.edu}
}

\maketitle

\begin{abstract}
Psychologists have demonstrated that pets have a positive impact on owners' happiness. For example, lonely people are often advised to have a dog or cat to quell their social isolation. Conventional psychological research methods of analyzing this phenomenon are mostly based on surveys or self-reported questionnaires, which are time-consuming and lack of scalability. Utilizing social media as an alternative and complimentary resource could potentially address both issues and provide different perspectives on this psychological investigation. In this paper, we propose a novel and effective approach that exploits social media to study the effect of pets on owners' happiness. The proposed framework includes three major components: 1) collecting user-level data from Instagram consisting of about 300,000 images from 2905 users; 2) constructing a convolutional neural network (CNN) for pets classification, and combined with timeline information, further identifying pet owners and the control group; 3) measuring the confidence score of happiness by detecting and analyzing selfie images. Furthermore, various factors of demographics are employed to analyze the fine-grained effects of pets on happiness. Our experimental results demonstrate the effectiveness of the proposed approach and we believe that this approach can be applied to other related domains as a large-scale, high-confidence methodology of user activity analysis through social media.

\end{abstract}


%
\IEEEpeerreviewmaketitle

\section{Introduction}
\label{intro}
The studies on analyzing the factors which can influence the happiness status and subjective well-being are of interest to researchers from a wide range of areas including psychology, economy, politics and healthcare. To quantify happiness and life satisfaction on a global scale, the Gross National Happiness (GNH) has been widely used for assessing the country-level well-being and happiness \cite{brooks2008gross,thinley2004values} and has motivated national happiness studies including the annual World Happiness Report launched by the United Nations \cite{helliwell2015world} since 2012. These measurements and reports on happiness are based on conventional methods including surveys, polls and self-report questionnaires, which are considered to be costly in terms of time and labor, and are challenging to maintain for long-term and large-scale studies. Moreover, participants involved may not always respond honestly to questions that make them feel uncomfortable, which results in potential bias in the research studies using such methods.

\begin{figure}[t]
  \centering
  \includegraphics[width=3.15in]{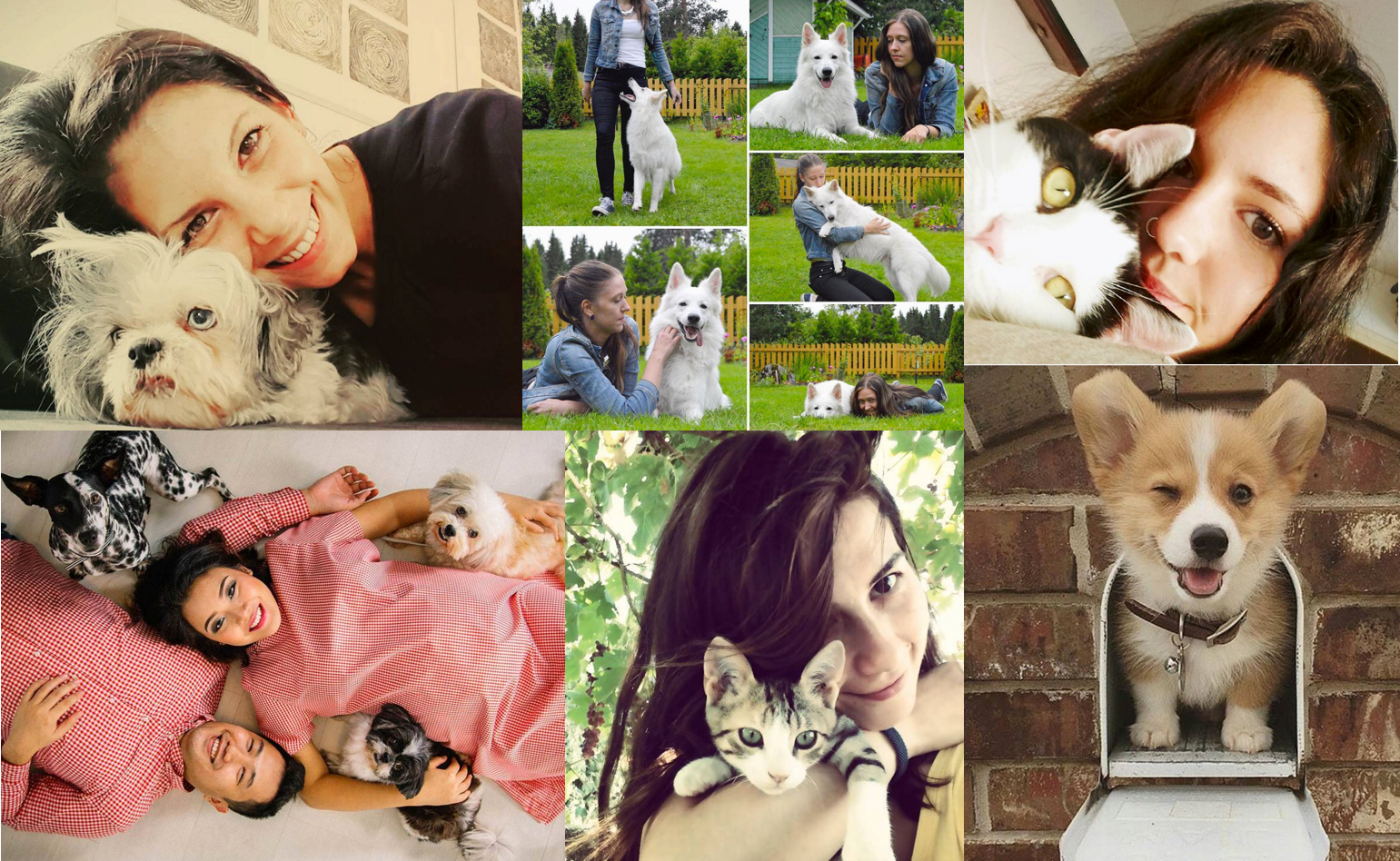}
  \caption{Examples of Pets Images From Instagram.}
  \label{pet_inst}
\end{figure}

With the explosive development of social media platforms such as Twitter, Facebook and Instagram where users tend to post texts and images to express their feelings and experience in their daily lives, there is an opportunity to take advantage of the massive and rich data generated by social media as a supplementary resource to the conventional methods for user behavioral and psychological studies. Compared with conventional methods, it is less time-consuming and labor-intensive to collect large-scale user level data from the social media platforms. Additionally, research based on social medias can be extended to a large scale and tracked for a longer time frame for long-term monitoring purpose. Moreover, social media presents an opportunity to observe the users in a natural, undisturbed state. In this paper, we aim to analyze whether pets have an impact on owners' happiness using social media. More specifically, we propose a computational approach to a fine-grained analysis on the pets effect on happiness through Instagram. After collecting images from 2905 Instagram users for six months (examples shown in Figure \ref{pet_inst}), we first employ a convolutional neural network (CNN) classifier combined with timeline analysis for pet owner classification. Next, we apply the state-of-the-art face engine \emph{Face++} for selfie image detection and user demographics inference for further analysis on pets effect on happiness. In summary, our contributions are as follows:
\begin{itemize}
	\item We propose a computational framework of a large-scale study on the effect of pet on happiness based on social media platform.
	\item We apply the state-of-the-art deep learning technique for pet owner classification and face analysis, and demonstrate the ability of our approach to capture users' psychological signals from social media.
	\item We propose an alternative approach to the conventional methods by utilizing social media, which can be also extended to study other user-level and macro-level behavioral and psychological related problems.
\end{itemize}

\section{Related Work}
\label{related}
The massive data generated from social media and its extensive usage makes social media a favorable platform for analyzing user behaviors and psychological states and provides a scalable and low-cost solution to such tasks. Related to our happiness study, sentiment analysis on textual contents, visual contents and multimedia contents has been addressed in previous works \cite{yuan2015sentiment,you2015joint}. However, sentiment analysis focuses more on the latent sentiments conveyed by the carriers such as images rather then inferring the state of users. In \cite{abdullah2015collective,hernandez2012mood}, computational metrics are proposed to measure happiness by collecting the intensity of smiles and has been validated against both text-based sentiment and self-reported happiness status \cite{abdullah2015collective}. Therefore we infer the happiness of Instagram users in our experiments following the same strategy. Another area of the related work is applying deep learning techniques to solving user-content related computer vision tasks: pet detection and face analysis. Deep neural networks such as CNN have shown great success in the tasks of object recognition \cite{krizhevsky2012imagenet,jia2014caffe}, as well as face analysis including face detection and demographics inference \cite{zhou2013extensive,fan2014learning}. We build the pet classifier based on CNN which is further combined with timeline analysis for pet owner identification, and apply the state-of-the-art face++ engine for face analysis for a fine-grained analysis on the pet effect on happiness.

\section{Data Collection}
\label{data}
\begin{figure}[h]
	\centering
	\includegraphics[width = 3.4in]{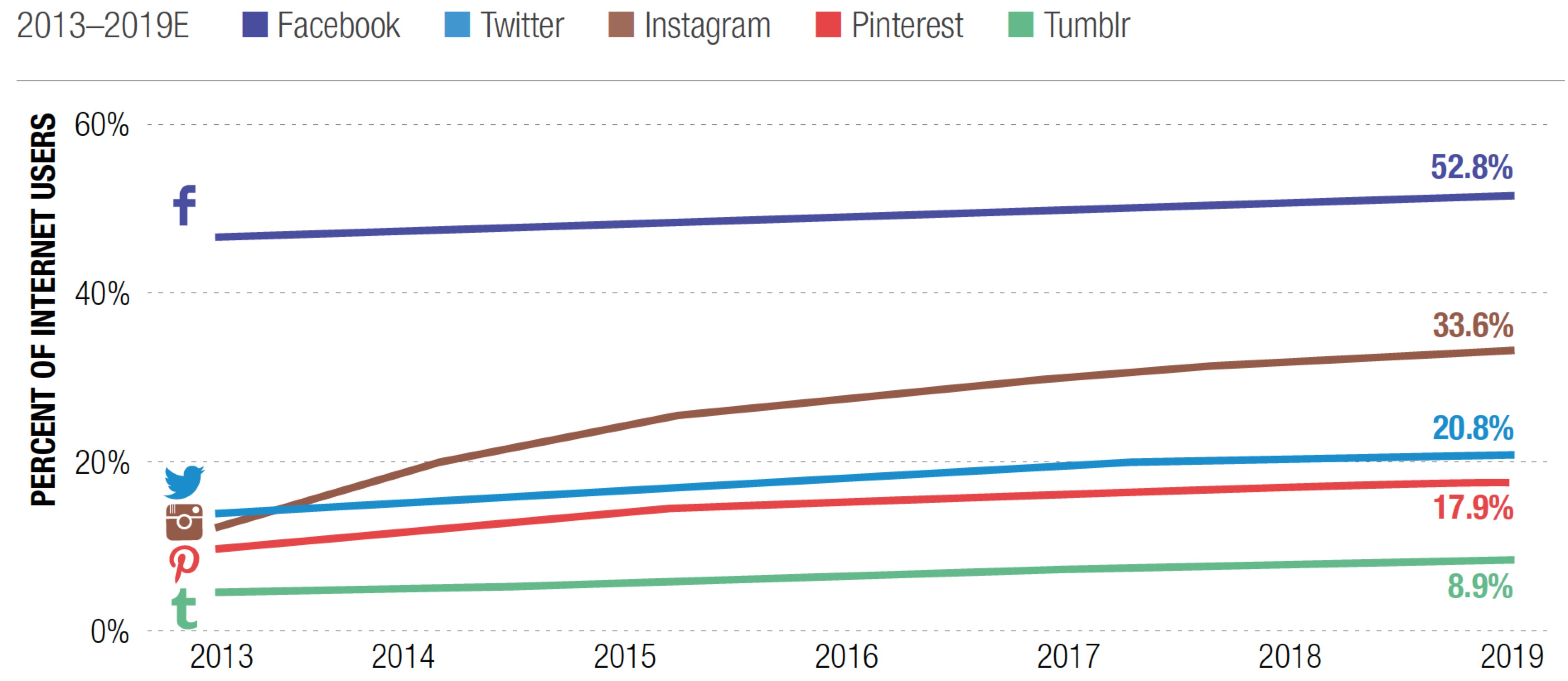}
	\caption{Predicted Penetration by Percent of Internet Users in the U.S. (Quoted from \cite{l2report}).}
	\label{instag}
\end{figure}

Social media has become the most popular platform for exposing and acquiring information between users. There are several features about Instagram which make it a good choice for analyzing the pet effect on happiness. Instagram has become one of the most important social media platform since 2014 as shown in Figure \ref{instag} \cite{l2report}, especially for teenagers. Compared with Instagram, Twitter is more focusing on information propagation such as spreading breaking news and interesting facts other than expressing the feelings of users. Although Facebook owns the largest user population, it is not mainly a platform for image and video sharing, and it has a more restrictive API usage which makes it difficult to extend the research based on Facebook to a large scale. 

On the other hand, Instagram is an image sharing-oriented social media that provides us with a good input resource to work on for image based analysis, where interactions between users are based on image sharing and users are more likely to share information about their personal and daily life experiences and express their feelings. The level of user engagement is also very high that about 49\% of the Instagram users are on the platform daily, and about 32\% of the users visit it several times per day\footnote{http://www.pewinternet.org/2015/01/09/frequency-of-social-media-use-2/}. Additionally, user demographics are needed for a fine-grained analysis but can hardly be collected directly from any social media platforms. With Instagram and a large number of available selfie images, we can apply the state-of-the-art face++ engine for user demographics inference. Instagram provides a more flexible API usage than Facebook, so that account information such as user self-tags and geo-location, user connections such as followers and followees, and more importantly the image posts including the metadata, can be collected. For the data collection, we first collected random image posts in December 2015 through the Streaming API. In this way we ensured the initial samples are randomly generated without any bias from the API. We then selected the users who have multiple selfies (no less than three) detected by the face++ engine and crawled all the images posted within the six months from June to December in 2015 for these users. In the end, we use the Instagram API to crawl 2905 users with selfie images and their image posts from June to December in 2015, where the images amount to a total number of more than 300,000. The user-level data is then processed by our pet classifier and face++ engine for further analysis.

\begin{figure*}[t]
	\centering
	\includegraphics[width = 6.5in]{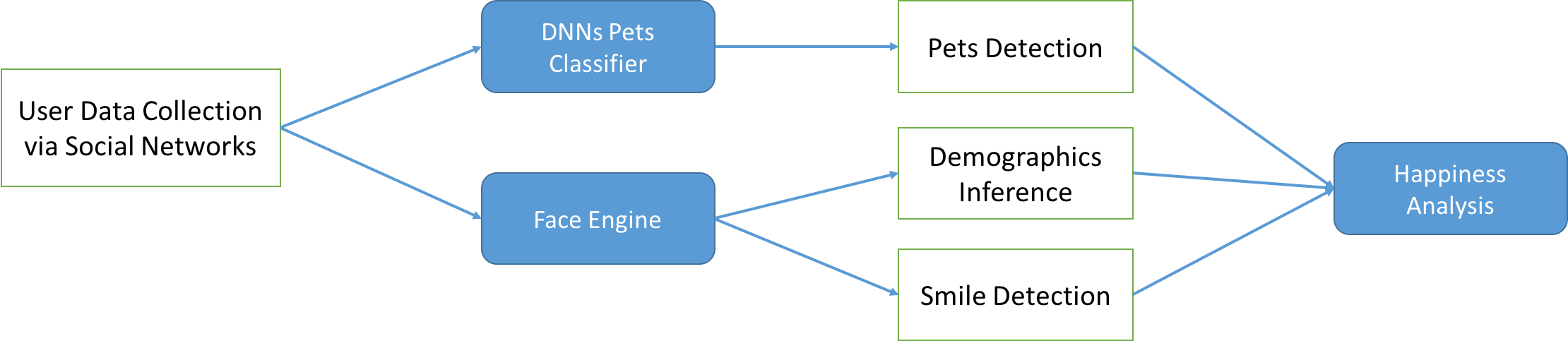}
	\caption{Proposed Framework for User Happiness Analysis.}
	\label{frame}
\end{figure*}

\section{Methodology}
\label{method}
Since Instagram does not require users to provide their real names and demographics including age, gender and race, we propose a large-scale image based computational analysis framework to study the pets effect on happiness. This can also be extended to analyzing general user behaviors via social media. Instagram contains very important self-representative information of its users among which are the selfie images. By applying the start-of-the-art face++ engine on selfie images, we are able to solve more complicated tasks beyond face detection and face recognition, including inferring user demographics such as age, gender and race, as proposed in our framework shown in Figure \ref{frame}. In order to determine whether the users own pets or not based on the images we collected, we build a convolutional neural network (CNN) for pet (cats and dogs only in this study) detection. Users without pets are identified for the control group for the later happiness analysis. Meanwhile, all the selfie images detected from within the six-month time frame are processed by the face++ engine again for user demographics inference and smile detection which are used subsequently for the purpose of the fine-grain happiness study.

\subsection{Pet Owner Classification}
\label{sec-pet}
\begin{figure}[!]
	\centering
	\includegraphics[width = 2.8in]{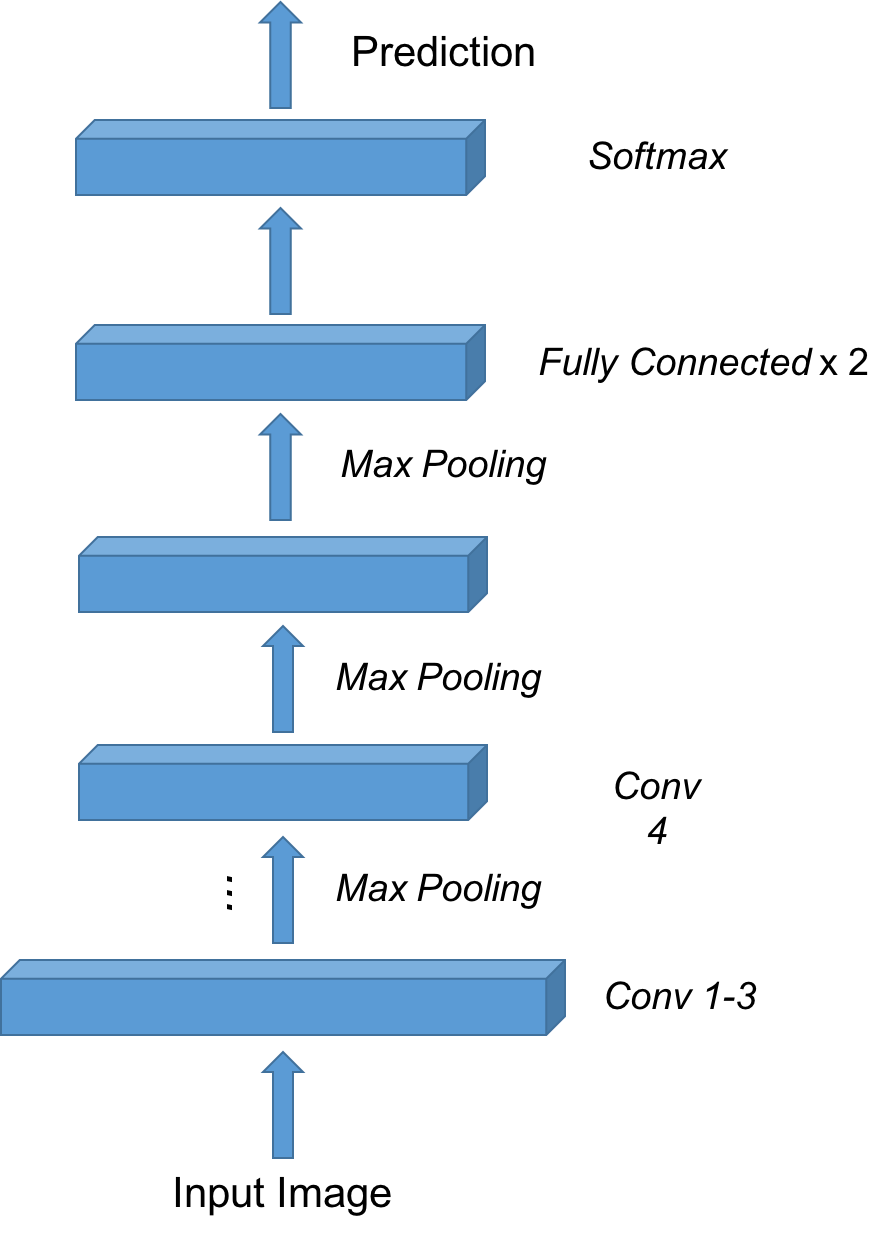}
	\caption{The CNN Classifier for Pet Detection \cite{krizhevsky2012imagenet}.}
	\label{imgnet}
\end{figure}

After collecting data for the selected users, the very first step of this study is to classify the users into two groups, one with pets as the experiment group and the other without pets as the control group. To this end, we build a CNN classifier to identify whether there is a pet in the image posts or not following the network proposed in \cite{krizhevsky2012imagenet} as show in Figure \ref{imgnet}. Considering that the most common pets in the United States are cats and dogs, we try to classify all the input images into three groups in this paper: cats, dogs and other.

\begin{figure}[h]
	\centering
	\includegraphics[width = 3.3in]{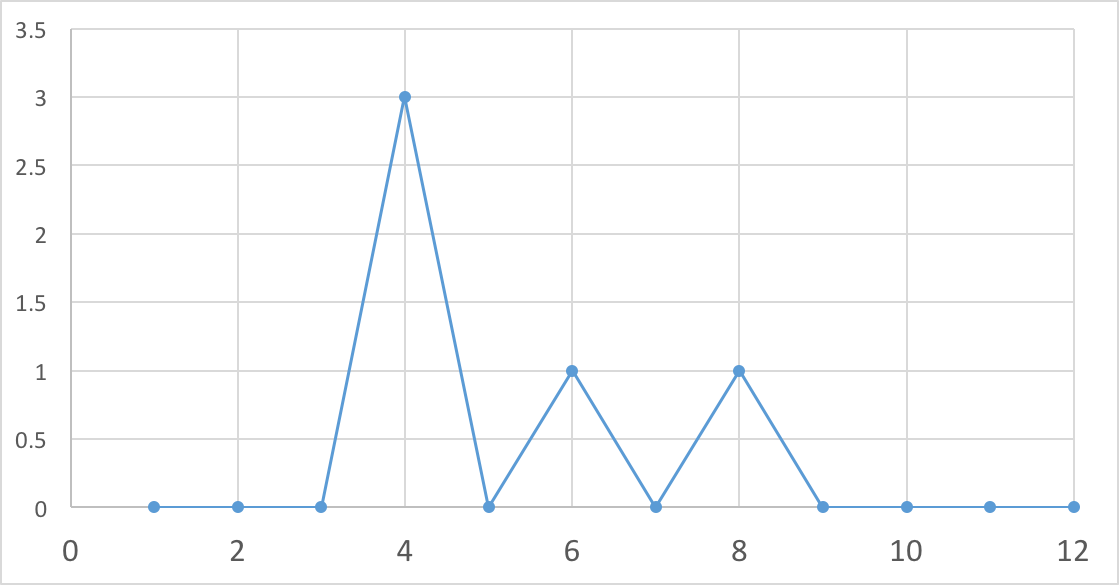}
	\caption{An Example of Pet Posts on a Pet Owner's Timeline.}
	\label{pet}
\end{figure}

In order to train this classifier, we basically use three datasets: 1) The Stanford Dogs dataset, which contains images of 120 dog breeds with a total number of 20,580 images, and has been built using images and annotation from ImageNet for the task of fine-grained image categorization \cite{KhoslaYaoJayadevaprakashFeiFei_FGVC2011}; 2) The CAT dataset, which contains 10,000 images of different cats where the images have a large variations in scale, pose and lighting \cite{cat2008}; and 3) The image posts extracted from random Instagram users, with 34,000 images in total, manually labeled as ``cats", ``dogs", or ``others" since the labeling test doesn't require additional training. The Stanford Dogs dataset and the CAT dataset are used for pre-training the CNN and the dataset we collected from Instagram is used for fine-tuning among which the category ``others" are used as negative samples in pet classification. Once we labeled all the images posted by the users, we further identify whether a user owns a pet or not. By examining the users whose image posts are classified as containing pets, we find that for some of the Instagram users, the pet images only appear once in their timeline, or contain different pets. These users are considered to be a ``pet lover" who tend to post images about pets, or post images of other people's pets. Therefore, we establish a criteria to distinguish pet owners from the others: we assume that if one user owns pets, the user may post the pet images of the same type of pets (i.e. only dog or cat) at different times throughout the timeline. On the other hand, if a person posts the pet images only within one short period of time (one week), this person is considered to have taken photos of pets in friends' homes or somewhere regardless of how many pet images have been posted. Classified images together with timeline information analysis are applied as a criterion to identify whether the user owns pets. Figure \ref{pet} is an example of the pet posts over the timeline of a classified pet owner, which shows multiple peaks of pets images being detected. 

\subsection{Selfie Detection, Demographics Inference and Happiness Analysis}
\begin{figure}[h]
	\centering
	\includegraphics[width = 3.3in]{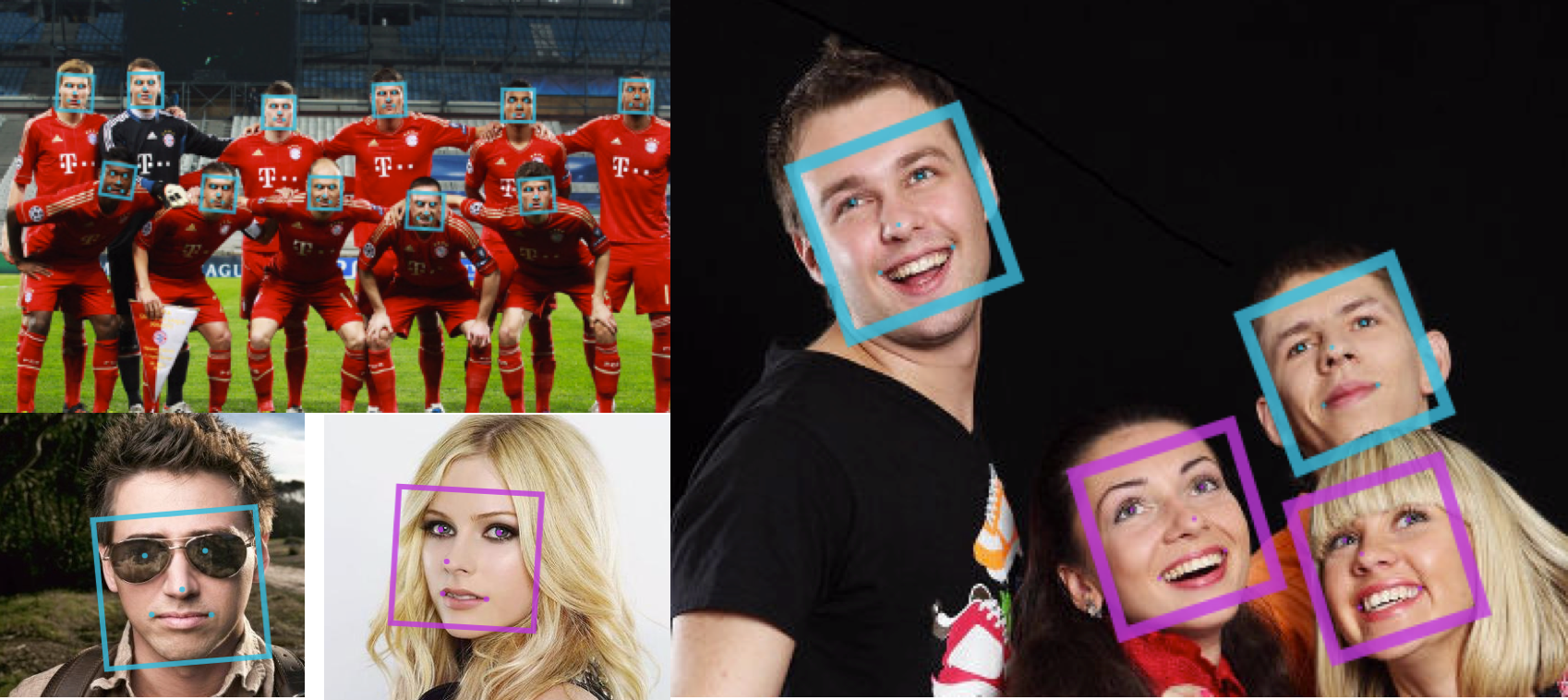}
	\caption{Examples of Face Detection from \emph{Face++}.}
	\label{face}
\end{figure}

For the tasks of selfie image detection, user demographics inference and happiness analysis, we propose to apply the state-of-the-art engine called \emph{Face++} \cite{zhou2013extensive}. \emph{Face++} is an open source face engine with both online API and offline SDK which provides services including face detection, face recognition and face analysis. The system is built with a CNN structure similar to the ImageNet structure as discussed in section \ref{sec-pet} \cite{krizhevsky2012imagenet}, where five convolutional layers with maxpooling connected with two fully connected layers and a softmax layer on top of them \cite{zhou2013extensive}. Since the performance of \emph{Face++} for face detection is very reliable, we use it for selfie image detection in this study. We apply post-processing on the images which are detected to have face by the \emph{Face++} engine such that images are classified as selfies if they only contain a single detected face whose size is larger than 10\% of the whole image calculated by the bounding box. Figure \ref{face} shows the examples of face detection\footnote{http://www.faceplusplus.com/demo-detect/}.

We assume that the selfie images detected belong to the user of the account which we want to analyze, and infer the demographics information (gender) and perform smile detection based on all the selfie images for each user. The gender classification by \emph{Face++} performs decently with a precision of 93\% and a recall of 94\% when tested on 1000 random selfie images from Instagram. In order to infer the happiness status of the users, we adopt the same methodology as \cite{abdullah2015collective} to compute the smile index based on the smile detection results returned from the face engine as in Equation \ref{smile-equ}. Let $HI_{t}$ denotes the happiness index for one user in a certain time frame $t$:
\begin{equation}
	\label{smile-equ}
	HI_{t} = \frac{\sum_{i\in I_{t}}s_{i}}{\left | I_{t} \right |}
\end{equation}
where $s_{i}$ is the confidence score of the smile detection result of a face image $i$ which belongs to the whole image set denoted by $I_{t}$ for the same user in time $t$. Please note that we apply smile detection and use the confidence score for all images detected as  containing faces other than limiting to selfie images \cite{abdullah2015collective}. For images with multiple faces detected, we apply the smile detection only on the biggest face in that image based on the assumption that it is the closest to the camera and the most likely to be the user.

\section{Experiments}
\subsection{Pet Detection Results}
Recent research on deep learning has shown great success in computer vision, where many CNN structures have achieved the state-of-the-art performance on computer vision problems such as object recognition. The network discussed in \cite{krizhevsky2012imagenet} was proposed to solve a multiclass classification problem (1000 classes), while in our case we have a much simpler case which only includes three classes. We pre-train the CNN proposed in Section \ref{sec-pet} and fine-tune the structure with 34,000 manually labeled Instagram images. We further test the classifier with another 1000 images randomly selected from our collected Instagram image dataset with manual labels as well. The confusion matrix is shown in Table \ref{conf}. Our application scenario is relatively simple which only involves three classes, the CNN classifier performs well and achieve an accuracy of over 96\% for each class. Additionally, the classification of the class ``other" (none-pet) is with a high recall value of 99\% which works well in our case since it indicates that only very few of the none-pet images are classified as pet images and thus would not adversely affect our judgment on whether the user we analyze is a pet owner or not. We then feed the classifier with all the images we collected from Instagram. Combined with the time stamp for each pet image, we can further identify each user as a pet owner and none-pet owner accordingly.
\begin{table}[h]
	\centering
	\caption{Confusion Matrix of the CNN Pet Classifier.}
	\label{conf}
	\begin{tabular}{|c|c|c|c|c|}
		\hline
		\multicolumn{2}{|c|}{\multirow{2}{*}{Confusion Matrix}} & \multicolumn{3}{c|}{Predicted} \\ \cline{3-5} 
		\multicolumn{2}{|c|}{} & Cat & Dog & Other \\ \hline
		\multirow{3}{*}{Actual} & Cat & 96.20\% & 1.90\% & 1.90\% \\ \cline{2-5} 
		& Dog & 0.80\% & 97.70\% & 1.50\% \\ \cline{2-5} 
		& Other & 0.40\% & 0.60\% & 99.00\% \\ \hline
	\end{tabular}
\end{table}

\subsection{User Classification and Happiness Analysis}
We use \emph{Face++} for selfie image detection and demographics inference since \emph{Face++} is the state-of-the-art open-access face engine on tasks of face detection and face analysis \cite{fan2014learning,zhou2013extensive}. We first use selfie detection to target users who have posted multiple selfies in a certain time frame and then apply face analysis on all the images we collected from these users within six months. In total we have 2905 users with selfie images. Face analysis from \emph{Face++} provides smile detection and demographics inference including age, gender and race. In the process of race classification, we find that the majority of users are white and only a few of them are none-white. The small numbers of users for other races make the data very sparse and not suitable for studying the relationship between race and happiness. Therefore with the demographic information we only analyze the relationship between gender and pet effect on happiness because the gender classification performs decently with an overall accuracy over 90\%. After pet owner classification and gender inference, we categorize the users into the following groups as shown in Table \ref{user}. There are about half males and half females in our dataset, and 33.8\% users are classified as pet owners among whom females have a slightly higher rate of being pet owners at 34.9\% while 32.6\% of males are pet owners. It is noteworthy that with such a large user base, we can adjust the proportion of demographics to match a targeted population. 

\begin{table}[h]
	\centering
	\caption{User Partition after Gender and Pet Classifications.}
	\label{user}
	\begin{tabular}{|c|c|c|c|}
		\hline
		& Have Pets & Don't Have Pets & Sum \\ \hline
		Male & 439 & 909 & 1348 \\ \hline
		Female & 543 & 1014 & 1557 \\ \hline
		Sum & 982 & 1923 & 2905 \\ \hline
	\end{tabular}
\end{table}

\begin{figure}[!]
	\centering
	\includegraphics[width = 2in]{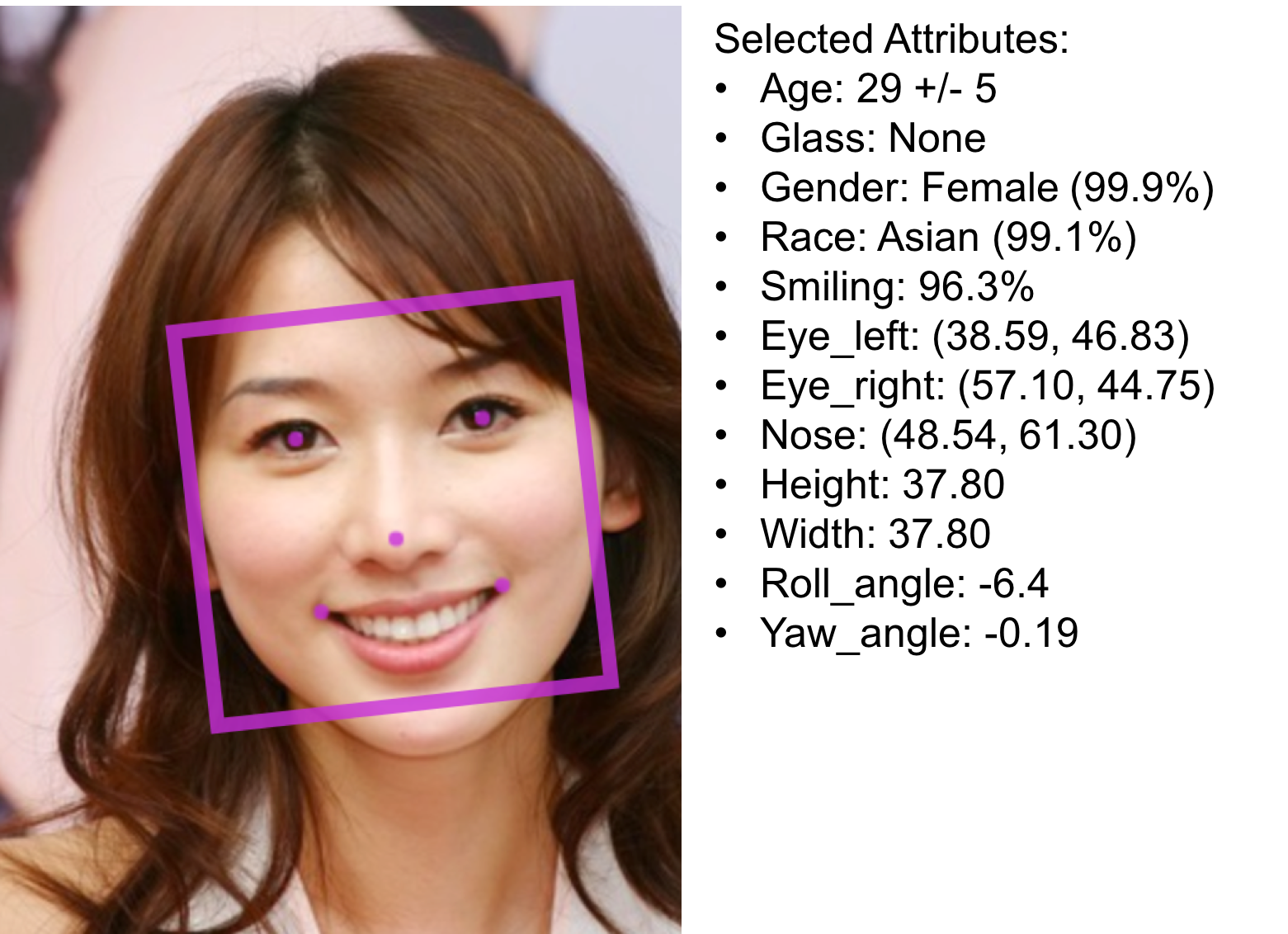}
	\caption{An Example of Face Analysis with Selected Attributes from \emph{Face++}.}
	\label{facetest}
\end{figure}

\begin{figure}[!]
	\centering
	\includegraphics[width = 3.4in]{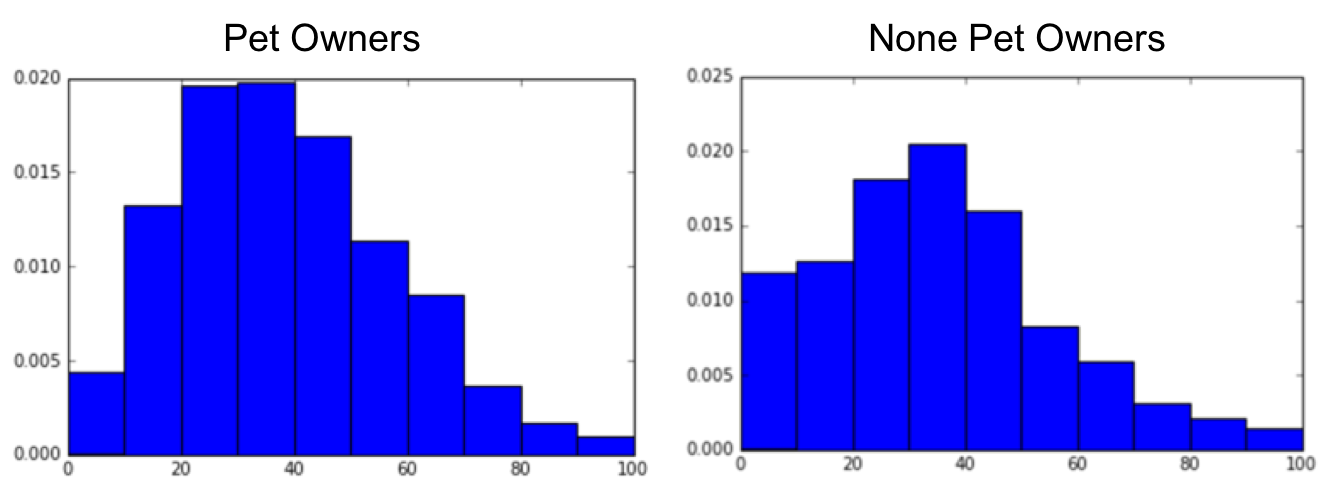}
	\caption{The Happiness Distribution for Pet Owners and None-Pet Owners ($\chi^{2}=84.27$, $p=0.0001$).}
	\label{fig-general}
\end{figure}
\begin{figure}[!]
	\begin{minipage}[!]{0.5\linewidth}
		\centering
		\includegraphics[width = 3.4in]{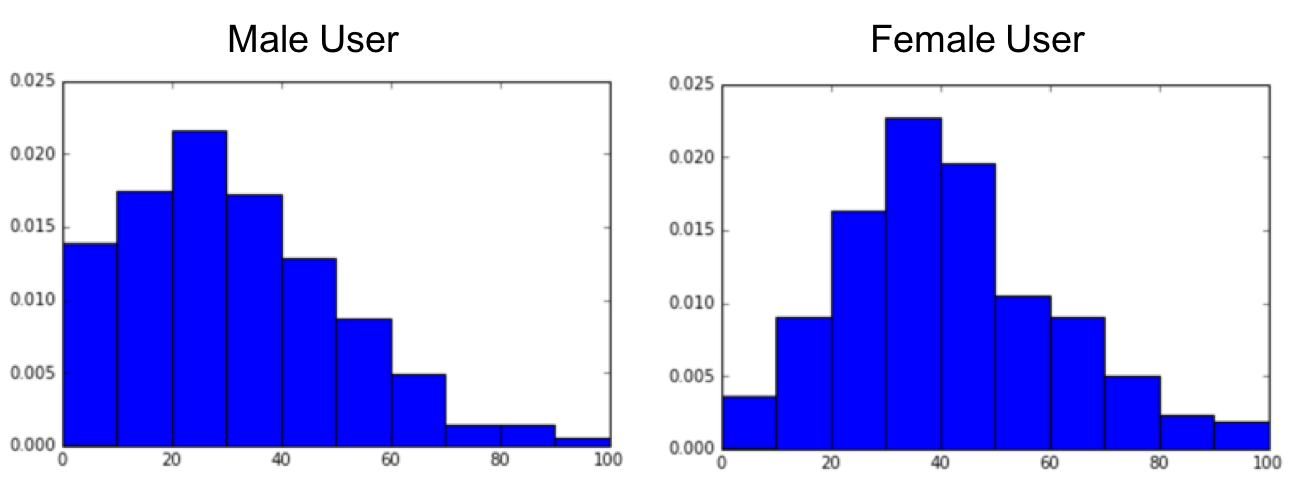}
	\end{minipage} \\
	
	\begin{minipage}[!]{0.5\linewidth}
		\centering
		\includegraphics[width = 3.4in]{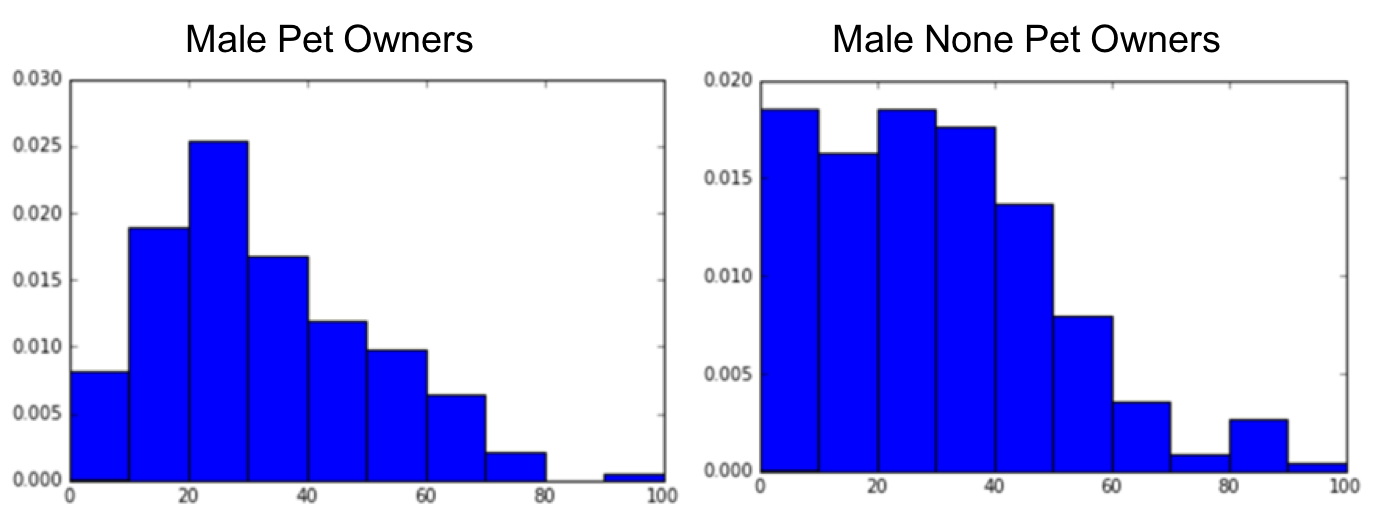}
	\end{minipage} \\
	\begin{minipage}[!]{0.5\linewidth}
		\centering
		\includegraphics[width = 3.4in]{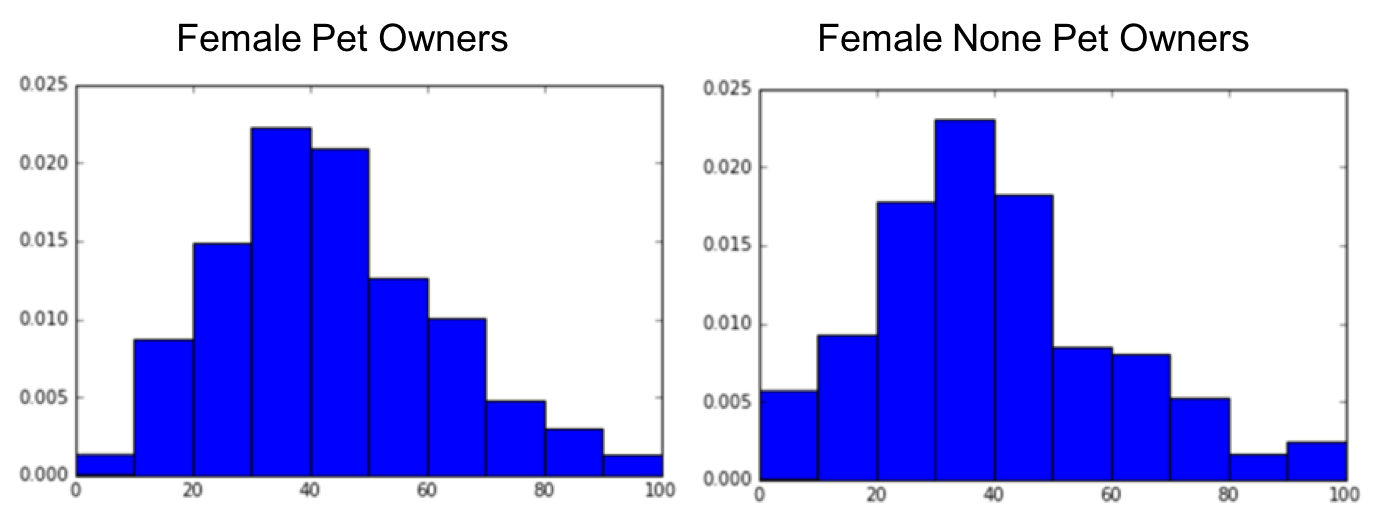}
	\end{minipage}
	\caption{Happiness Analysis on Male and Female Pet and None-Pet Owners: (a). The happiness index distribution of male (left) and female users (right); (b). male pet owners (left) and male none pet owners (right); (c). female pet owners (left) and female none pet owners (right). }
	\label{gender}
\end{figure}

An example of the returned demographic inference and smile detection is shown in Figure \ref{facetest}. Given the confidence score of smile detection on all the images with faces for each user, we compute the happiness index based on Equation \ref{smile-equ}. We then analyze the distribution of happiness index for different categories of users as presented in Figure \ref{fig-general} and \ref{gender}. These histograms have been normalized in order to see the general distribution. The x-axis is the happiness index score ranging from 0 to 100 while the y-axis is the portion of users in a certain happiness score range normalized by the total number of users under the specific category. A chi-square test of independence demonstrates the two distributions are significantly independent with $p=0.0001$. Generally, users with pets have less portion in the low happiness score range of 0-20, and have more portion of users in the high happiness range of 50 to 80. Such difference in the happiness score distributions indicates a general happier state of users who have pets over those none-pet owners. Further examination on the distributions of the happiness index for male pet owners and none-pet owners in Figure \ref{gender} (b) shows a similar trend to the general distribution of pet owner and none-pet owner in Figure \ref{fig-general} where male pet owners show a happier status than the male none-pet owners. On the other hand, comparison between female pet owners and none-pet owners does not show a difference as apparent as the male users in Figure \ref{gender} (c). However, the female pet owners shows a larger portion of users fall in range 50-70 and smaller portion of range 0-20 than the female none-pet owner which indicates the pet effect on happiness. It is surprising to conclude from the results that male users are more sensitive to pet effect than female users because some psychology studies show that females are more sensitive and more expressive about feelings. One  explanation for this finding is that, as shown in Figure \ref{gender} (a), generally female users show a much happier status than the male users which results in a less impact on the happiness index caused by the pet effect. Analysis on races shows little insights due to our lack of users detected of different races. Another key factor is that users on Instagram tend to post positive images on their accounts as a social effect observation, which can explain the similar large portion of mid-range (20-40) happiness score users with and without pet effect.

\section{Conclusions and Future Work}
In this study, we propose a data-driven approach of using Instagram to studying the effect of pets on happiness. A CNN-based pet classifiers and timeline analysis are combined for pet owner classification, and the state of the art \emph{Face++} engine is applied for face detection and face analysis. In contrast to conventional methods such as surveys, our proposed approach represents a scalable and low-cost solution to user behavioral and psychological study. The happiness analysis can be combined with sentiment analysis to monitor user’s emotional status over time through social media. Moreover, the proposed approach is an example of using social networks for users psychological status inference and can be further extended to risky behavior modeling such as underage smoking, healthy diet monitoring, and so on. For behavior modeling, more sophisticated classifiers are expected for incorporating text content such as posts and hashtags, in conjunction with image classifiers in order to achieve robust results.

\section*{Acknowledgment}

We gratefully acknowledge the support of New York State through the Goergen Institute for Data Science. 


\bibliographystyle{plain}
\bibliography{ref}

\end{document}